\documentclass{aa_new}    
\usepackage{graphicx}

\newcommand{\like}{{\cal L}}

\newcommand{\Ho}{H_{\rm o}}
\newcommand{\Omb}{\Omega_{\rm b}}
\newcommand{\etab}{\eta_{10}}

\newcommand{\lamo}{\lambda_{\rm o}}
\newcommand{\Omk}{\Omega_\kappa}
\newcommand{\OmT}{\Omega_{\rm tot}}

\begin{document}

   \title{Cosmology from Cosmic Microwave Background and Galaxy Clusters}

        \titlerunning{Cosmology from CMB and Galaxy Clusters} 

   \author{M.~Douspis$^{1,2}$, A.~Blanchard$^{1}$,  R.~Sadat$^1$, J.G.~Bartlett$^1$,  M.~Le~Dour$^1$} 
   \authorrunning{M. Douspis et al.}

   \offprints{\email{douspis@astro.ox.ac.uk}}

   \institute{$^1$ Observatoire Midi-Pyr\'en\'ees,
              14, ave. E. Belin,
              31400 Toulouse, FRANCE \\
              Unit\'e associ\'ee au CNRS, UMR 5572
             ({\tt http://webast.obs-mip.fr})\\    
       	      $^2$ Astrophysics, Nuclear and Astrophysics Laboratory,
                   Keble Road,
		   Oxford, OX1 3RH,
	           UNITED KINGDOM\\
             }

   \date{May 2001}

   \abstract{
We present the results of  analysis of constraints on cosmological parameters
from cosmic microwave background  (CMB) alone and in combination with
galaxy cluster baryon fraction assuming  inflation--generated adiabatic 
scalar fluctuations. The CMB constraints are obtained using our likelihood 
approximation method (Bartlett et al., 2000, Douspis et al., 2001). 
In the  present analysis we use the new data coming from MAXIMA and
BOOMERanG  balloon borne experiments, the first results of the DASI
interferometer together with the COBE/DMR data.  The quality  of these
independent  data sets implies that the  $C_\ell$ are rather well
known, and allow reliable  constraints. We found that the constraints
in the $\Omega-H_0$ plane are very tightened, favouring a flat
Universe,  that the  index of the primordial fluctuations  is very
close to one, that the primordial baryon  density is now in good
agreement with primordial nucleosynthesis. Nevertheless degeneracies
between  several parameters  still exist, and for instance  the
constraint on the cosmological constant or the Hubble constant are
very weak, preferred values being low. A way to break these
degeneracies is to ``cross-constrain'' the parameters by combining
with constraints   from other independent data. We  use the baryon
fraction determination from X--ray clusters of galaxies as an
additional constraint and show that the combined analysis leads to
strong constraints on all cosmological parameters. Using a high baryon
fraction ($\sim 15\%$ for $h = 0.5$) we found   a rather low Hubble 
constant, values around $80$ km/s/Mpc being ruled out. Using a recent 
and low baryon fraction estimation ($\sim 10\%$ for $h = 0.5$) we found  a
preferred model with a low Hubble constant and a high  density content
($\Omega_m$), an  Einstein--de Sitter model  being only weakly ruled
out.
      }

\maketitle


\section{Introduction}
The determination of cosmological parameters is a central goal of modern
cosmology. The measurements of the angular spectrum of the fluctuations in the 
CMB is one of the most promising techniques. Indeed the first measurements of 
fluctuations at the degree scales (Saskatoon, Netterfield et al., 1997)
has allowed to put the first constraints
 on cosmological parameters (Lineweaver et al. 1997; Hancock et al 1997).
 Probably the most spectacular result was that 
the open models could be reasonably  excluded, while flat models were
favoured (Lineweaver \& Barbosa; 1998b), essentially because of the
location of the first peak. 
 This result has been brilliantly confirmed by the first BOOMERanG and MAXIMA 
measurements (Hanany et al., 2000,  de Bernardis et al., 2000). The 
data are  now  becoming of high quality and 
the  detailed features expected in so-called inflationary models are now 
becoming apparent (the famous peaks) from independent measurements
(BOOMERanG, MAXIMA, DASI). 
Alternative scenarios like topological 
defects models are almost entirely ruled out. The possibility of a partial 
contribution remains open (Bouchet et al., 2000) and will be
difficult to  entirely rule out,
although inflationary models are clearly preferred. In this paper we analyse 
the constraints that can be set from the recent measurements obtained by 
BOOMERanG, MAXIMA, DASI combined with the COBE/DMR results.  

However, it has been realised that combination of different methods to
constraint  cosmological parameters was very welcome, not only because
of  the gain in precision, but mainly because most of cosmological
tests  present  degeneracies and may constraint rather well only 
specific combinations  of cosmological parameters but not each
parameter individually. Combinations of 
constraints or measurements, often  implemented as prior on some parameters in
likelihood analysis,  are currently used to obtain accurate values for
the cosmological  parameters. In this paper, we use as a first
additional constraint  the observed baryon fraction in clusters. This
quantity has the advantage of not specifying one
fundamental cosmological parameter  in  a direct way.  Nevertheless, we 
obtained  contours which are significantly tightened for all the
parameters, breaking all the degeneracies.  This
allows us to investigate constraints on cosmological parameters without 
introducing any prior from  measurements based on standard candles hypothesis.
In this paper we use $H_{\rm 0} = 100h$ km/s/Mpc.

\begin{figure}

\begin{center}
\resizebox{\hsize}{!}{\includegraphics[angle=0,totalheight=8.4cm,
        width=8.cm]{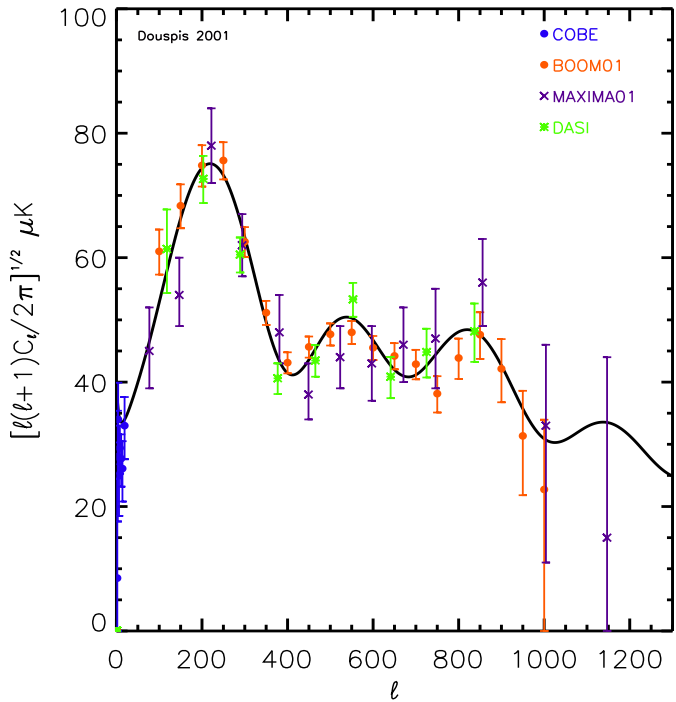}}
\end{center}
\caption{\label{fig_powplot}The power plane: measured flat--band power 
estimates (COBE, BOOMERanG, MAXIMA and DASI) and  
our best fit model (by approximate maximum likelihood): $(\Ho,\OmT,\lamo,\Omb h^2,n,Q)=(30\; $km/s/Mpc,$
1.3,0.1,0.019,0.91,23.0\; \mu$K).
}

\end{figure}    

\section{Constraints from CMB alone}

\subsection{CMB analysis}

During the last decade (since the COBE/DMR results) different measurements
of CMB anisotropies at different scales have been used to constrain the 
cosmology of our Universe (Hancock et al., 1997 , Lineweaver et al.,
1997, Dodelson and Knox, 2000, Tegmark and 
Zaldarriaga 2000ab, Ledour et al., 2000, Jaffe et al., 2001 and others).
 The last 
experiments (second generation) are sensitive to different
scales. This gives  homogeneous sets of data  from a few degrees to a few
arcmins, which cover two or three peaks in the power spectrum with good
accuracy.
We then expect the constraints to be stronger, especially if several
experiments are consistent.

Our method to derive parameters from CMB data has been presented in Bartlett
 et al., 2000 (BDBL), Douspis et al., 2001 (DBBL) and used in Le Dour et al.,
 2000 (LDBB)\footnote{see http://webast.ast.obs-mip.fr/cosmo/CMB for complementary informations}. We use the maximisation technique rather than the marginalisation
 obtained by integrating over some parameters. Both techniques are equivalent 
when the probabilities are Gaussian and the model is linear in the 
parameters. This is far from being the case however when dealing with 
likelihood on the $C_\ell$. The degeneracies imply that  likelihood surfaces 
 are sheets in a multidimensional  space. In the presence of such a complex 
structure in the likelihood space, the marginalisation has the inconvenient 
that the model corresponding to the preferred parameters may actually lie
 outside of the actual allowed region! We present our results by means of 
contours in 
 two dimensional parameter space. This allows us to some extent to identify 
complex structure in the multidimensional parameter space, that might be hidden
 in  the likelihood on one parameter.
Goodness of fit has to be evaluated before addressing parameter estimation.
The technique to derive the goodness of fit in our 
method is detailed in Douspis et al. 2001 (DBB). The basics are to use
flat--band estimates as published (see table 2 of LDBB) and apply our
 likelihood approximation instead of a non appropriate $\chi^2$ minimisation.
Our method remains as simple as a $\chi^2$ fit but is less biased  as shown
 in DBBL. In this paper, we use  the most recent results of
 MAXIMA (Lee et al., 2001), BOOMERanG (Netterfield  et al., 2001) and DASI
 (Halverson et al., 2001). We also use the COBE/DMR data (Tegmark and
 Hamilton, 1997) and refer to the data set as CBDM. 
 We also upgraded our grid of 
models, allowing closed models and refining the step and range of the baryon 
content ($\Omega_bh^2$), the total density ($\Omega_{tot}$) and the
 spectral index  
($n$). The parameter space explored is given in table 1.

\begin{table}[h]
\begin{center}
\begin{tabular}{|c|c|c|c|c|c|c|}
\hline
 & $\Ho$ km/s/Mpc&$\OmT$&$\lamo$&$\etab$&$n$&$Q\; \mu$K\\
\hline
\hline
Min. & 20  & 0.7   & 0.0   & 2.78  & 0.70  & 10.0 \\
\hline
Max. & 100 & 2.0   & 1.0   & 11.94 & 1.30  & 30.0 \\
\hline
step & 10  & 0.1   & 0.1   & 0.83  & 0.03  & 1.0  \\
\hline
\hline
\end{tabular}
\end{center}
\caption{Parameter space explored:} 
\vspace{-0.3cm}
$\OmT \equiv 1-\Omk$, where $\Omk$ is the curvature parameter \\
$\lamo \equiv \Lambda/3$ \\
$\etab \equiv$ (baryon number density)/(photon number density)\\
\hspace*{0.9cm} $\times 10^{10}$ \hspace*{0.2cm} (Note: $\Omb h^2 
= 0.00366\etab$)\\
$n\equiv$ primeval spectral index\\
$Q\equiv \sqrt{(5/4\pi) C_2}$ \\
\end{table}

We do not consider reionisation or gravitational waves nor neutrino 
contributions to the matter density $\Omega_m$. Previous analyses
showed that these parameters  do not play an important role.
The present analysis  has been done with all the information available in the
literature. Unfortunately, some information is still missing (like the 
window functions);  some effects have  thus not been taken into account but 
our previous investigation of these questions indicates that this lack of information is not critical.

 The likelihood of each model
is calculated as described in LDBB, and the
best model is found by maximising the
likelihood function over the explored space.
The contours are defined in the full six--dimensional
space with values 
of $\Delta\log(\like)=1, 4, 9$ (dashed, in red),
corresponding to the 1, 2 and 3 $\sigma$ contours 
for  Gaussian distributions 
when projected onto one of the axes. 
Identically   the 1, 2 and 3 $\sigma$ contours in the two parameters space
are defined by 
$\Delta\log(\like)=2.3, 6.17, 11.8$ (filled, in blue), for 
Gaussian distribution.  Since the likelihood is not actually Gaussian, 
the confidence percentages associated with our 
contours are actually not known; the technique is
however standard practice.

We also would like to emphasise the effect of priors in the analysis. As it 
has been noticed in the recent literature (Jaffe et al., 2001, Balbi et al.,
 2000, Lange et al., 2000), the final results on the preferred parameters are 
very sensitive to priors. Most of the recent work set $H_0$ as a ``ghost'' 
parameter, using physical densities 
($\Omega_bh^2,\; \Omega h^2,\;\Omega_{dm}h^2, ...$) and deriving $H_0$ by:
$$h=\sqrt{\frac{\omega_{dm}+\omega_b}{1-\Omega_k-\Omega_\Lambda}}.$$ Then the
 analysis proceed by putting a prior on $H_0$ like $H_0 > 45$. The 
representation in one-dimensional plots does not allow  to know
exactly what is the effect of such technique. We will show in section
\ref{cont_cmb}  that 
such a prior puts more severe constraints on $\Omega_{total}$ as it
 is  expected. 

The study of the goodness of  fit of ``best'' models  
is detailed in DBB. We apply the same technique on the new data and
 the  best model plotted in figure \ref{fig_powplot}. The goodness of fit
 expressed in terms of generalised $\chi^2$ is $\chi^2_{gen}=45.0/43$ where 
$43=49-6$ is the number of degrees of freedom corresponding to the number 
of experimental points minus the number of investigated parameters.
This good value of the goodness of fit allows us to consider confidence 
intervals on cosmological parameters.

\subsection{Constraining the cosmological parameters \label{cont_cmb}}

\begin{figure*}
\begin{center}
\resizebox{\hsize}{!}{\includegraphics[angle=0,totalheight=11cm,
        width=12cm]{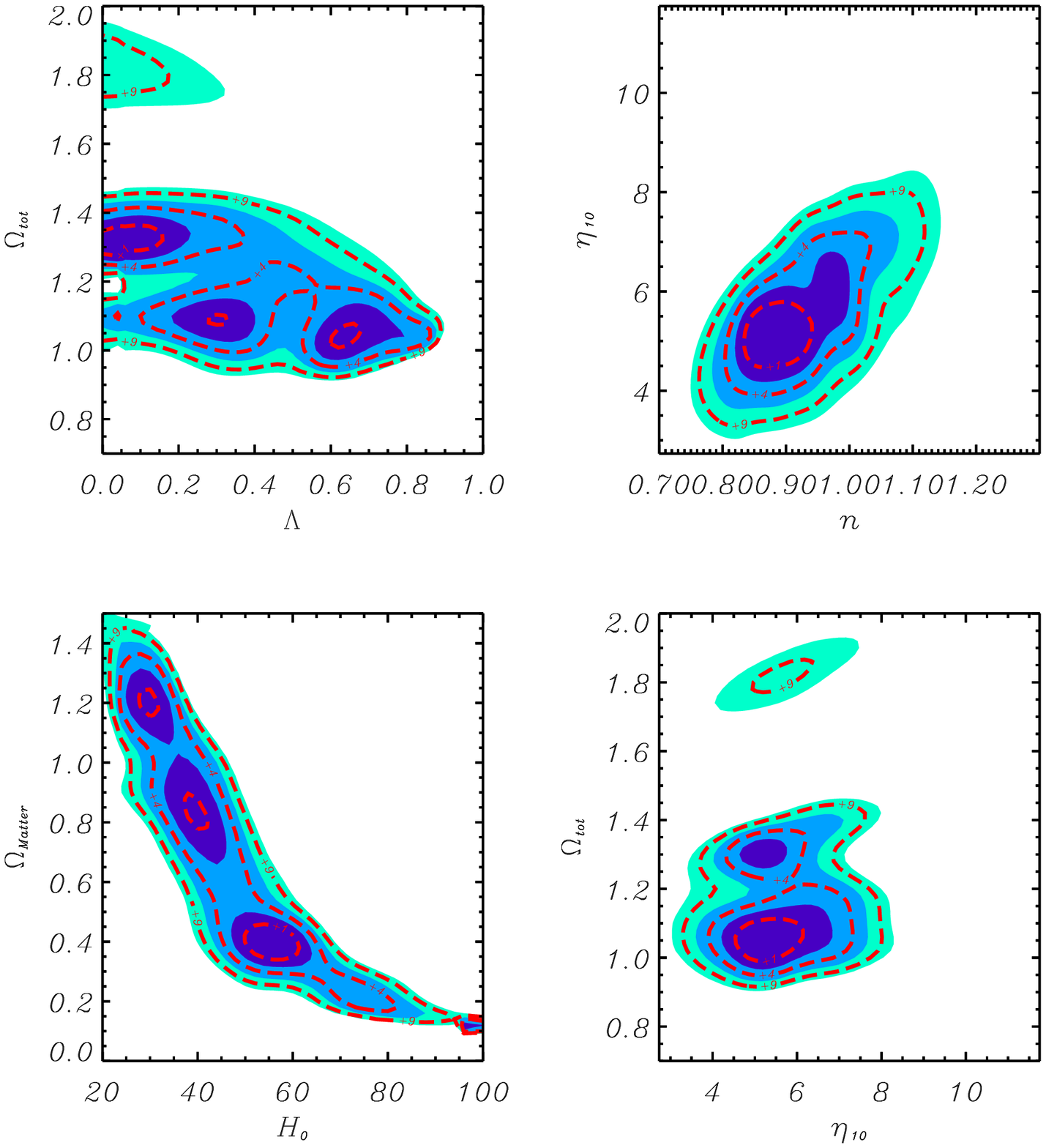}\includegraphics[angle=0,totalheight=11cm,width=12cm]{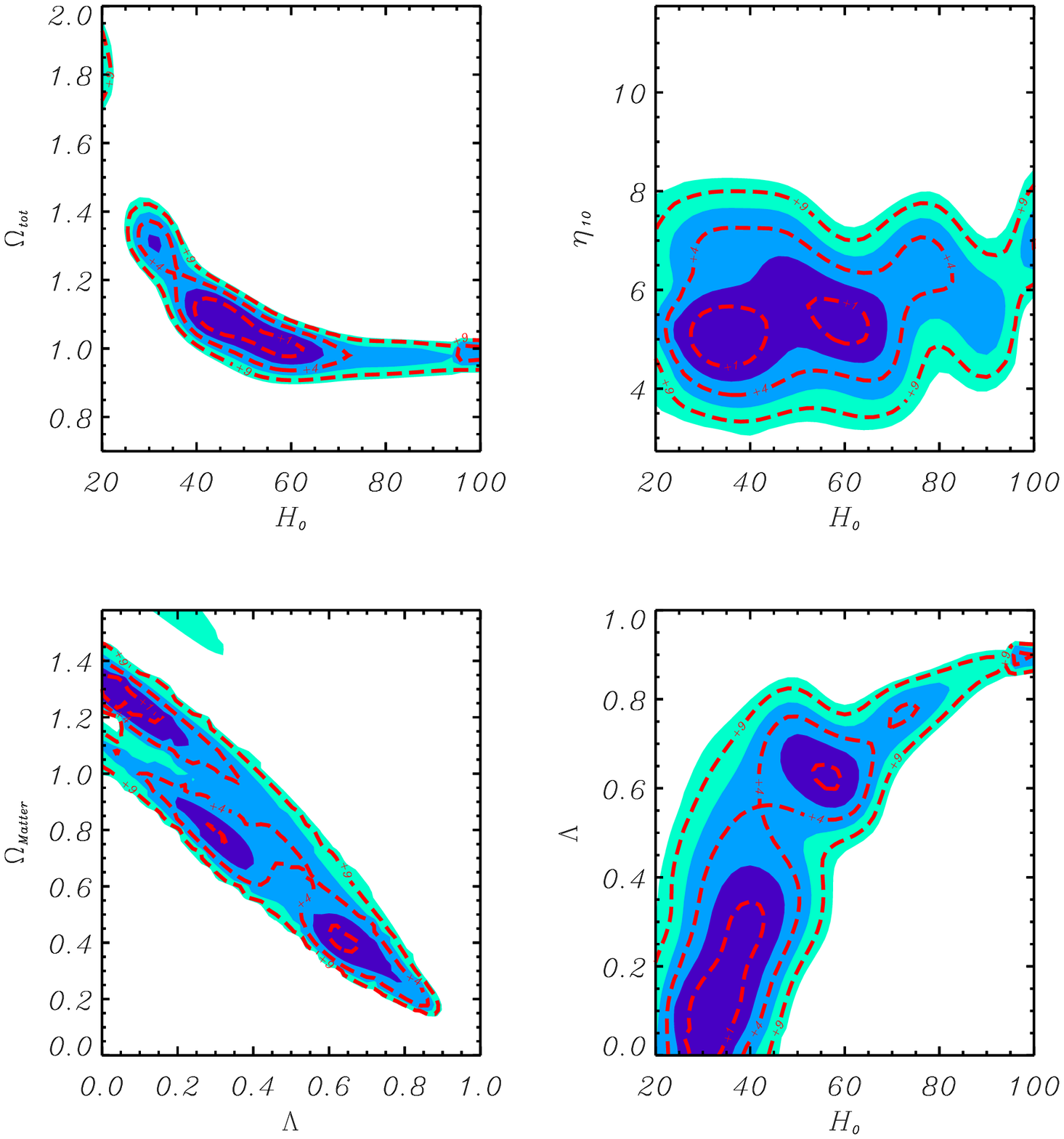}}
\end{center}
\caption{\label{fig_CMB}Different contour plots from the analysis of set 
{\bf CBDM}. The dashed red line define the 68, 95 and 99\% GCL when
projected on the axis. The blue filled contours are the corresponding
confidence intervals in 2 dimension.}
\end{figure*}

The first Doppler peak is one of the main feature of the CMB angular
 power spectrum. It is clear now that different experiments indicate both 
a rise in the power at intermediate scale  and  a fall--off at
smaller scales (higher $\ell$). The position of the first acoustic peak is
 strongly related to the curvature of the Universe. A statistical 
analysis of
the CMB data should then give some strong constraint on the total density 
(or curvature) of the Universe. The confidence intervals on the 
$\Omega_{total}$ parameter in different combinations 
($(\Omega_{total},\Lambda), (\Omega_{total},H_{\rm
 0}), (\Omega_{total},\eta_{10})$) 
 are shown as 2-D contour plots in figure \ref{fig_CMB}.
The $\Omega_{total}-H_0$ plane reveals the stronger constraint that CMB fluctuations
 lead to: the contours are almost reduced to a line in this plane.
 The main implication  is the exclusion of almost all
open cosmologies at high  confidence level. Preferred models 
 are closed ($\Omega_{total}\sim 1.3$) and   relative maxima
 are present at 
lower value ($\Omega_{total}\sim 1.05$). 
The models corresponding to 
the latter are in good agreement 
with a flat Universe and
the inflationary scenario prediction. Within the 95\% Gaussian confidence
 level (95\% GCL hereafter), and allowing $H_0$ to be in then range
 [20,100] we find:   $0.95 < \Omega_{total} < 1.4$. 
Given our 2-D representation, it is easy to see that the closest models have 
the lowest value of $H_0$ (cf. figure \ref{fig_CMB}). This is a supplementary
 information that the 1-D reduction is unable to show.

The fourth plot of figures \ref{fig_CMB} presents the results in 
the $\eta_{10}-\Omega_{total}$ plane. The constraints on
$\eta_{10}$ are becoming   tight 
with the present day data set compared to 
 analysis of first generation experiments (LDBB).  They point toward
a medium  value of the baryon content which  
is well in agreement with the BBN and light elements determination of $\eta_{10}$, contrarily to  analysis based on first  BOOMERanG and MAXIMA results 
(Tegmark \& Zaldarriaga, 2000, Douspis, 2000, Jaffe et al., 2001). The 
change in the preferred value of $\Omega_bh^2$  is due to the improved
measurements of the second and third peaks. 
In spring 2000, BOOMERanG and MAXIMA experiments measured the power at
scales corresponding to the second acoustic peak. The latter appears to be
lower (relative to the first one) than expected in a model with a baryon 
content compatible with BBN predictions ($\Omega_bh^2 \sim 0.02$, Burles 
et al 2001). The relative heigh between the two peaks favours high values of
$\eta_{10}$ corresponding to $\Omega_bh^2 \sim 0.03$. In spring 2001,
 BOOMERanG, DASI and MAXIMA teams are publishing new analysis showing 
the possible existence of a third acoustic peak  with
the same height as the second. This disfavours high baryon models
which predict a higher third peak  (see figure 1 of LDBB). 
The actual value in accordance with our analysis is $4.0 <
\eta_{10} < 7.0$ at  95\% GCL and almost independently of the value of
$H_0$.

The spectral index of primordial fluctuations, $n$ is also a well constrained
 parameter. We found  $0.81 < n < 1.02$ at 95\% GCL. This is
in  agreement with the inflationary predictions.

The remaining parameters are almost  not constrained.
We can see in figure \ref{fig_CMB} the different degeneracies  between
parameters. 
There is no constraint on $H_0$  from the CMB. And we can also see that 
$\Omega_\Lambda$ can take almost any value: $0.0 < \Omega_\Lambda <
0.9$ at 95\% GCL. Another contour plot of figure \ref{fig_CMB} shows a tight
degeneracy between  
$\Omega_{total}$ and $H_0$.  
There again, it is obvious that the lower $H_0$, the higher $\Omega_{total}$.
Imposing $H_0 > $ 45 km/s/Mpc results in a tight constraint on
$\Omega_{total}$ : $0.9 < \Omega_{total}< 1.2 $ (99\% GCL). Finally,
the second plot of figure \ref{fig_CMB} points out the anticorrelation 
 of the pair ($n,\eta_{10}$).  Increasing $n$ (lowering $Q$) will increase the
 height of the second peak relative to the first. At the opposite, increasing 
$\eta_{10}$ will decrease the second peak. 

It is clear that the CMB alone does not directly constrain very
tightly $\Omega_{total}$ without  further restriction (or prior) on
$H_0$. Nevertheless it is important to keep in  mind that the CMB
impose a very narrow region in the  
$\Omega_{total}-H_0$ plane. Other parameters are not so well constrained. 
However, the baryon content is now becoming rather well constrained, 
in agreement with nucleosynthesis and light elements abundances and the 
primordial index is slightly below 1, close to the value predicted by
inflation.  
It is therefore a fundamental success of Cosmology that the CMB alone does 
allow such tight constraints. However, it is not possible yet to infer robust 
constraint on other fundamental parameters like $H_0$, $ \Omega_\Lambda$ or $ 
\Omega_m$. It is therefore vital to have other means to measure these 
fundamental parameters.

\section{Combined analysis}

Given the fact that CMB implies degeneracies between the fundamental
cosmological parameters $H_0$, $ \Omega_\Lambda$
and $\Omega_m$, it is interesting to look for combination with constraint that
 do not measure directly any of these parameters. Primordial nucleosynthesis  
is one of this possibility; we have checked that this does not change 
much the constraints established in the above section. An other possibility 
that we investigate below is to use the baryon fraction inferred from cluster 
observations.

\subsection{Baryon fraction in  clusters}

Clusters  provide  fundamental observations for
cosmology. Historically, they provided the first evidence for the
existence of dark matter. Furthermore they now provide us with the best
evidence that this dark matter is non baryonic.  Indeed, clusters are the only
structure for which the total mass, baryon mass and stellar mass can
be evaluated simultaneously in a reliable way. Cosmological
applications of clusters have been revived with new fundamental tests
which provide global  constraint on the cosmological parameters: the
baryon fraction in clusters trace reliably the  ratio between  baryons
and the total mass in the Universe: 
$$
f_b = \Gamma \frac{\Omega_b}{\Omega_0}
 $$
where $\Gamma$ has been estimated from numerical simulations and is close to $0.92$ in the
outer part of clusters (Frenk et al., 1999). Provided that $\Omega_b$
can be estimated from primordial nucleosynthesis and the observed
abundance of light elements, this argument can be used to estimate a
robust upper limit on $\Omega_0$ (White et al., 1993). In this paper
we combine CMB and the sole baryon fraction constraints. This is interesting
because the  baryon fraction does not constrain by itself any of the
cosmological  parameters, so that in principle it 
could leave unchanged the   conclusions  obtained from the CMB alone.  

\subsection{Combining constraints}

There are some dispersions among values published for the gas fraction
in clusters, which mainly reflect the difference in the mass estimator
used.  Using mass estimators based on numerical simulations a median
baryon   fraction of 
$0.048h^{-3/2} + 0.014\; (\pm 10\%)$ was recently found by Roussel et
al., 2000, consistently with previous investigations (Arnaud \&
Evrard, 1999). More recently the standard value has been revisited 
by Sadat  and Blanchard, 2001,  showing that the actual baryon
fraction could actually be lower.
They concluded that  a baryon fraction as  low as $0.031h^{-3/2} +
0.012\; (\pm 10\%)$ is still  consistent with the X--ray  data on
clusters.
 
\subsubsection{The high baryon fraction case}

\begin{figure}
\begin{center}
\resizebox{\hsize}{!}{\includegraphics[angle=0,totalheight=11cm,
        width=12cm]{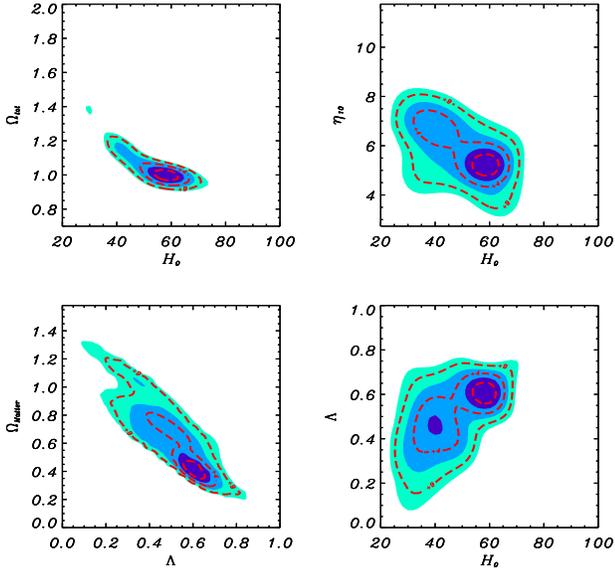}}
\end{center}
\caption{\label{fig_cmbhfb}Combined analysis of CMB and high baryon
fraction. See figure \ref{fig_CMB} for the definition of contours.}
\end{figure}

As we have mentioned previously, we have chosen to investigate the
combination of 
the baryon fraction $f_b$ with the CMB data. The baryon fraction is an
interesting case because its value can be measured in a rather direct
way through observations of X-ray clusters provided the assumption 
that all the baryons are actually seen (assuming no large quantity of
dark baryons). It is worth noticing that without  an additional
constraint on $\Omega_b$, there is no trivial reason why the 
additional  information on  $f_b$ should further constrain the various
cosmological parameters. However, it happens that the additional
constraint from the baryon fraction restricts significantly the
various contours (see figure \ref{fig_cmbhfb}). This is very clear from
the $H_0-\eta_{10}$ plane: the area of the contours is much
reduced, leading to a preferred model which is around $H_0 \sim 60 $
km/s/Mpc and $\eta_{10}\sim 5$. The other two-dimensional plots also
reveal that  all the contours are significantly reduced: actually most
of the one sigma contours are of the order of the grid size so that
best model and one sigma ranges  should be used with caution. It is
highly interesting that the cosmological model is now very well
constrained: 
$\Omega_{\lambda} \sim 0.6$,   $\Omega_{tot} \sim 1.$ (implying
$\Omega_{m} \sim 0.4$,  $H_0 \sim 60 $ km/s/Mpc, $\eta_{10}\sim 5$,
$n \sim 0.85$). As we mentioned the allowed range have to be
interpreted with caution,  but we notice that the preferred value of
the Hubble constant is rather low and that higher values like
80 km/s/Mpc  lie uncomfortably outside of the preferred region.

\subsubsection{The low baryon fraction case}

\begin{figure}
\begin{center}
\resizebox{\hsize}{!}{\includegraphics[angle=0,totalheight=11cm,
        width=12cm]{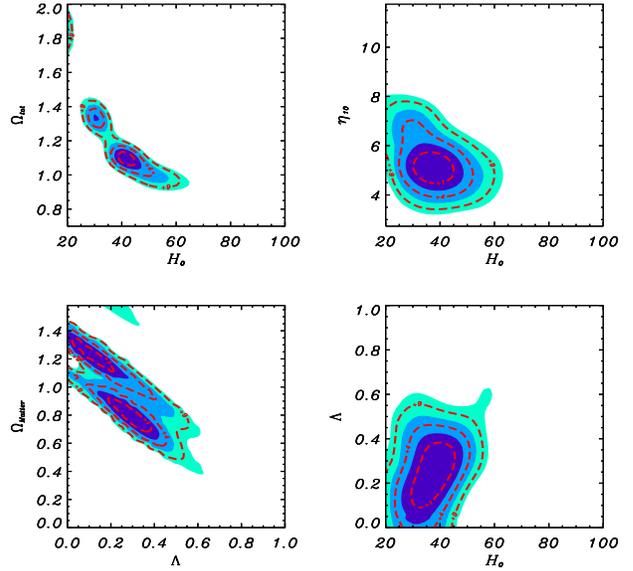}}
\end{center}
\caption{\label{fig_cmblfb}Combined analysis of CMB and low baryon
fraction. See figure \ref{fig_CMB} for the definition of contours.}
\end{figure}

As mentioned above, a baryon fraction
significantly lower than current estimates   is still consistent with, and
actually preferred by existing data on X-ray clusters. Using this low
value for the baryon fraction leads similarly to much restricted
contours. However the latter  differ from the previous ones
 (see figure \ref{fig_cmblfb}): the 
preferred value for $\eta_{10}$ is again close to 5., the preferred value of 
the Hubble constant is now very low ($H_0 \sim 40 $ km/s/Mpc). Although
such low  
values have been argued  from time to time (Bartlett et al., 1995), it is worth
 noticing that values  as $H_0 \sim 60 $ km/s/Mpc are still at the
edge of the 3  
sigma  contour and for this reason cannot be entirely ruled out. The preferred
 model is $\Omega_{\lambda} \sim 0.3$,   $\Omega_{tot} \sim 1.1$ (implying and
 $\Omega_{m} \sim 0.8$,  $H_0 \sim 40 $ km/s/Mpc, $\eta_{10}\sim 5.$, $n \sim 
0.85$). 

Finally, we also investigated the additional constraint that can be inferred 
from the evolution of the abundance of X-ray clusters: the evolution of the 
abundance of X-ray clusters is known to be a powerful constraint on the matter
 density of the universe (Oukbir and Blanchard, 1992). The measurement of the 
temperature of a flux-limited sample of  X-ray clusters has allowed the 
determination of the abundance of clusters at $z\sim 0.3-0.4$ (Henry, 1997; 
Henry, 2000). Using the updated estimates of local abundance of X--ray
clusters  a significant level of evolution is found, indicating  a high 
density Universe (Blanchard et al., 2000).

The  constraint provided by the addition of 
the evolution of cluster abundance is essentially a constraint on
$\Omega_m$.  Therefore it is rather natural   that this additional
constraint by itself will not change  much the previous contours as
the value  found by Blanchard et al., 2000, $\Omega_{m} \sim 0.8$, is
almost identical  to what is provided by the previous analysis.  We
can anticipate that using analysis leading to lower $\Omega_{m}$ with
the higher baryon fraction will not change much the  contours
presented in the previous section.

\section{Conclusions}
The new CMB data provide impressive evidence in favour of the standard
inflationary picture for structure formation with non baryonic dark
matter. They also brilliantly confirm previous evidence for a nearly
flat Universe. In this paper,  we have actually shown that the data
provide a very 
narrow allowed region in the   $H_0-\Omega_{tot}$ plane. However, it
is somewhat frustrating that despite the high quality of the data,
degeneracies among the fundamental cosmological parameters
($\Omega_{tot}, \Omega_{\lambda}$ and $H_0$) cannot be broken and that
constraints on other parameters are relatively weak.  It should also be
realised that given the dependency on the power spectrum shape and
the possible contribution at some level  from topological defects, it
will be extremely difficult to obtain entirely reliable constraints on
these parameters from CMB alone. Actually, even with the precision
anticipated for the Planck mission, the degeneracy problem may not be
solved. It is therefore vital to develop new tests that provide
additional constraints. Supernova Hubble diagram  provide such a test
which has be widely advertised. However, the previous use of the
Hubble diagram has  always been polluted by systematic effects, and
therefore we believe that a coherent picture will be firmly
established only from further evidence.  Clusters offer such
additional possibility. We have therefore investigated the complementary
information which come from  clusters. The most interesting
constraint has been found, rather surprisingly,  to be the baryon
fraction. Indeed with this single additional constraint we found that
nearly all the cosmological parameters are well specified. Using a
classical  value for the baryon fraction of $\sim 15 \%$ for $h = 0.5$
we obtain as the best model: $\Omega_{\lambda} \sim 0.6$ ,
$\Omega_{tot} \sim 1.$,  $H_0 \sim 60 $  km/s/Mpc, $\eta_{10}\sim 5.$,
$n \sim 0.85$. Using a more recent result from Sadat and  Blanchard,
2001,   who infer a lower baryon fraction when several systematics
 are  corrected for, we found that the best model is
$\Omega_{\lambda} \sim 0.3$,   $\Omega_{tot} \sim 1.1$ ,  $H_0 \sim 40
$ km/s/Mpc, $\eta_{10}\sim 5.$, $n \sim 0.85$ implying a high density
parameter  $\Omega_{m} \sim 0.8$ consistent with the determination
from cluster abundance evolution by  Blanchard et al., 2000. Although
the Einstein--de Sitter model is at the edge of the 3 sigma contour, it
is probably  premature to rule it out on this basis. It is remarkable
that in both  analyses we found that the preferred primordial baryon
content is narrowly  constrained and consistent with primordial
nucleosynthesis.  Identically the primordial index is found to be of
the order of $0.85$ in both cases. 
 


\end{document}